\documentclass[12pt,a4paper ]{article}
\usepackage[cp1251]{inputenc}
\usepackage[russian]{babel}

\usepackage{graphicx}
\input epsf

\newcommand{\frat}[2]{\frac{\textstyle #1}{\textstyle #2}}
\newcommand{\vf}[1]{\mbox{\boldmath $#1$}}

\begin{document}

\begin{center}
{\Large \bf  Properties of lightest mesons at finite temperature
and quark/baryon chemical potential in instanton model of QCD
vacuum}\\ \vspace{0.5cm} S.V. Molodtsov$^{1,2}$, A.N.
Sissakian$^{1}$, A.S. Sorin$^{1}$, G.M. Zinovjev$^{3}$ \\
\vspace{0.5cm} {\small $^1$Joint Institute for Nuclear Research,
RU-141980, Dubna, Moscow region, RUSSIA.}\\ \vspace{0.5cm} {\small
$^2$Institute of Theoretical and Experimental Physics, RU-117259,
Moscow, RUSSIA.}\\ \vspace{0.5cm} {\small $^3$Bogolyubov Institute
for Theoretical Physics, UA-03680, Kiev-143, UKRAINE}
\end{center}
\vspace{0.5cm}

%\begin{abstract}

\begin{center}
\begin{tabular}{p{16cm}}
{\small{The thermal and quark/baryon chemical potential
dependences of quark condensate and masses of $\pi$- and
$\sigma$-mesons are studied in the instanton model of the QCD
vacuum in precritical region. The impact of phonon-like
excitations of instanton liquid on the characteristics of
$\sigma$-meson in such an environment is also examined.}}
\end{tabular}
\end{center}

%\end{abstract}
\vspace{0.5cm}

%PACS: 11.15 Kc, 12.38-Aw

%\baselineskip 0.9cm

Nowadays with the common belief that the hadron world and its symmetries are driven by QCD
our understanding of the lightest mesons origin, at least from the hadron side, is tightly related to the
spontaneous breakdown of chiral symmetry inherent to the corresponding Lagrangian. Then the pions appear
as the massless Goldstone bosons while the scalars acquire a non-zero vacuum expectation values reflecting
the sophisticated structure of QCD vacuum. At the level of QCD fundamental fields we have phenomenologically
learnt that this symmetry breaking is dominated by the non-zero values of quark condensates \cite{SVZ} and
the scalar meson attributes result from the dynamical generation of light quark masses. However, at rather
high temperature $T$, as the lattice QCD studies have taught \cite{1}, the phase transition restoring the
chiral symmetry occurs (and a quark chiral condensate is just its order parameter) together with the transition
deconfining the quarks and gluons and another transition of colour superconductivity (characterized by a diquark
condensate) at high quark/baryonic chemical potential $\mu$. Clearly, the last phase transition is scarcely
feasible in the terrestrial conditions and is relevant rather in the astrophysics observations of compact stars.
As to two others they are intensively explored in ultrarelativistic heavy ion collisions \cite{1a}.

The properties of various phases of hot and dense matter are dependent on the interplay between the competing
processes of quark-quark and quark-antiquark pairing (influenced by the QCD vacuum) and, as expected, could be
clarified by studying the QCD phase diagram on ($\mu$---$T$)-plane in the lattice simulations (albeit a very
idea is relevant for the infinite matter and in thermodynamic equilibrium). The first results of lattice phase
diagram studies were quite schematic but,at the same time, indicative enough because they generated very
important question about the structure of phases around the deconfinement critical temperature $T_c$, i.e. at
$|T - T_{c}|\ll T_c$. Today as become especially clear when it was found that $J/\psi$ charmonium survives at
rather high temperatures beyond $T_c$ \cite{1b}. In fact, this result has been confirmed by the RHIC experimetal
observations and has led to the concept of strongly coupled quark-gluon plasma \cite{1c} revealing the
existence of enormous number of bound states at $T > T_c$ in deconfined phase.

Phenomenological analysis of strongly interacting matter under the extreme conditions is complicated by the
comparatively low credibility of dynamical hadron models (linear and nonlinear sigma-models,
the Nambu--Jona-Lasinio model and its modificatios etc.) on the ($\mu$ -- $T$)-plane. Their predictions are
not likely to be reliable at approaching the critical line of phase transition restorating the chiral symmetry.
At small values of $\mu$ and $T$ the parameters of those models are tuned in such a way to reproduce some
experimental data and to obey some general constraints rooted in QCD. However, extending the models to the
phase transition regions leads sometimes to conflicting conclusions (see, for example, the recent discussion
in \cite{3}) At the same time the analysis of precritical regions could be entirely informative which was
demonstrated in \cite{3a} where the properties of lightest mesons in hot and dense environment making use the
virial expansion were estimated in chiral perturbation theory. In the present paper similar questions are
discussed within the instanton liquid model of the QCD vacuum \cite{4}.

The quark generating functional for stochastic ensemble of gluon
configurations being defined as (the sign of double averaging
means the corresponding average over the gluon fields) $${\cal
Z}_\psi~\simeq~\int D\psi^\dagger D\psi~~\langle\langle~
e^{S(\psi,\psi^\dagger,A)}~\rangle\rangle_A~,$$
at nonzero
temperature and finite chemical potential is estimated by the
saddle point method and could be presented (up to the details
inessential for our consideration here) with an auxiliary
integration over saddle point parameter $\lambda$ (we are dealing
here with $SU(3)$ gauge group and quarks of two flavours $N_f=2$)
\cite{4a} in the following form
\begin{eqnarray}
\label{1} &&{\cal Z}_\psi\simeq\int d\lambda~\int D\psi^\dagger
D\psi~\exp\left\{
n~V_4\left(\ln\frat{n\bar\rho^4}{\lambda}-1\right)\right\}
\times\nonumber\\
%[-.2cm]
%\\[-.25cm]
&&\times\exp\left\{T\sum_{l=-\infty}^{\infty} \int \frat{d {\vf
k}}{(2\pi)^3}~\sum_{f=1}^{2}\psi^\dagger_{f}(k) ~(-\hat
k-i\hat\mu)~\psi_{f}(k)+{\cal L}_{int}\right\}~,\\
%\nonumber\\
&&{\cal L}_{int}=i\lambda~(\psi^{\dagger L}_1~L_1~\psi_1^{L})
(\psi^{\dagger L}_2~L_2~\psi_2^{L})+ i\lambda~(\psi^{\dagger
R}_1~R_1~\psi_1^{R}) (\psi^{\dagger
R}_2~R_2~\psi_2^{R})~.\nonumber
\end{eqnarray}
Here $\psi_f^{T}=(\psi_f^{R},\psi_f^{L}),~f=1,2$ are the quark
fields composed by the spinors of fixed chirality
$\psi_f^{L,R}=P_\pm~\psi_f,~P_\pm=\frat{1\pm\gamma_5}{2}$, $n$ is
the instanton liquid density, $\bar\rho$ is the mean size of
instanton liquid pseudoparticles, $V_4=T L^3$ denotes the 4-volume
of system and $\mu_\nu=({\vf 0},\mu)$. Working in the Euclidean
space we are summing up over $l$ by introducing the Matsubara
frequencies $k_4(l)=2\pi T\left(l+\frat{1}{2}\right)$ with the
convention $\Delta k_4 = T$. Summing up and integrating over
spatial momentum components in Eq.(\ref{1}) will further be marked
as $\sum\!\!\!\!\!\int$. Then the Lagrangian ${\cal L}_{int}$ of
4-fermion quark interaction is given with the chiral components as
$$(\psi^{\dagger L}_f~L_f~\psi_f^{L})= \int \frat{dp_f
dq_f}{(2\pi)^8}~\psi^{\dagger L}_{f\alpha_f i_f}(p_f)~ L_{\alpha_f
i_f}^{\beta_f j_f}(p_f,q_f;\mu)~\psi_f^{L\beta_f j_f}(q_f)~,$$ and
for the right hand fields one should change $L \to R$.

 Currently,
the corresponding effective Lagrangians for the $L$ ш $R$
components are set in terms of the quark zero modes. At zero
values of temperature and chemical potential the zero modes
approach \cite{5}, being combined with the mean field
approximation \cite{6}, is providing us with very reasonable
phenomenological results. But at finite temperature the
superposition of (anti-)calorons which are the periodic solutions
of the Yang-Mills equations \cite{7} looks much more relevant to
be used as the saturating configuration of suitable generating
functional. At high temperature, meanwhile, the chromoelectrical
part of the gluon field is expected to develop the mass term
$\frat{m_{el}^2}{2}~A_4^{2}$ in the effective Lagrangian (the
corresponding term in one-loop approximation has been already
found out in \cite{8}). It implies the saturating configurations
should be mainly rooted in the chromomagnetic sector. (In the
framework of instanton QCD vacuum model the alternative scenario
has also been developed with assuming the formation of
instanton-anti-instanton molecules \cite{9}.) The recent lattice
studies of correlators of the topological configurations
corroborated \cite{10} the chromomagnetic character of gauge
fields beyond the critical temperature.

Unfortunately, even the simplest caloron zero mode (in the momentum space) at zero value of quark chemical
potential $\mu$ may not be presented by the special functions and, hence, is impractical for the analytical
exploration. It is the major reason for us to develop here the approximate procedure in which we substitute the
explicit expression of zero mode $\psi[A(T,\mu);T,\mu]$ for its simplified form $\psi[A(0,\mu);0,\mu]$ and
imply the corresponding corrections could be calculated perturbatively. Such an oversimplification results in
the $L_{\alpha_f i_f}^{\beta_f j_f}$ kernels which are defined by the zero modes (the solutions of the Dirac
equations with a chemical potential $\mu$) in the form of $h_i$-functions as regards
\begin{eqnarray}
h_4(k_4,k;\mu)&=&\frat{\pi}{4 k}
\{(k-\mu-ik_4)[(2k_4+i\mu)f^{-}_{1}+i(k-\mu-ik_4)f^{-}_2]+\nonumber\\
&+&(k+\mu+ik_4)[(2k_4+i\mu)f^{+}_{1}-i(k+\mu+ik_4)f^{+}_2]\}~,\nonumber
\end{eqnarray}
\begin{eqnarray}
h_i(k_4,k;\mu)&=&\frat{\pi~k_i}{4 k^2}
\left\{(2k-\mu)(k-\mu-ik_4)f^{-}_{1}+(2k+\mu)(k+\mu+ik_4)f^{+}_1+
\right.\nonumber\\
&+&\left[2(k-\mu)(k-\mu-ik_4)-\frat{1}{k}(\mu+ik_4)[k_4^{2}+(k-\mu)^2]
\right]f^{-}_{2}+\nonumber\\ &+&\left.
\left[2(k+\mu)(k+\mu+ik_4)+\frat{1}{k}(\mu+ik_4)[k_4^{2}+(k+\mu)^2]
\right]f^{+}_{2}\right\}~,\nonumber
\end{eqnarray}
with $i=1,\dots,3$ and $k=|{\vf k}|$, if one considers the spatial
components of 4-vector $k_\nu$, and $$f_1^{\pm}=
\frat{I_1(z^\pm)K_0(z^\pm)-I_0(z^\pm)K_1(z^\pm)} {z^\pm}~,\\
f_2^{\pm}=\frat{I_1(z^\pm)K_1(z^\pm)}{z^2_{\pm}}~,~~
z^\pm=\frat{\bar\rho}{2}\sqrt{k_4^{2}+(k\pm\mu)^2}~, $$ with
$I_i,~K_i~(i=0,1)$ as the modified Bessel functions. Introducing
in the same time the scalar function $h(k_4,k;\mu)$ which is
related to the three-dimensional components as
$h_i(k_4,k;\mu)=h(k_4,k;\mu)~\frat{k_i}{k}~,i=1,2,3$ (when it does
not mislead we omit the arguments of $h_i$ functions at all) we
have then
$$L_{\alpha i}^{\beta j}(p,q;\mu)=S^{i k}(p;\mu)
\epsilon^{k l}~U^{\alpha}_{l}~U^{\dagger\sigma}_{\beta}
\epsilon^{\sigma n}S^{+}_{n j}(q;-\mu)~,$$
with
$S(p;\mu)=(p+i\mu)^{-}~h^{+}(p;\mu)~,~~
S^+(p;-\mu)=\stackrel{*}h\!\!{^{-}}(p;-\mu)(p+i\mu)^{+}$,
moreover, for the conjugated function it is valid
$\stackrel{*}h_\mu(p;-\mu)=h_\mu(p;\mu)$, and $\epsilon$ is
entirely antisymmetric matrix $\epsilon_{12}=-\epsilon_{21}=1$.
Here $p^\pm$ and similar symbols are used for the 4-vectors
spanned on the matrices $\tau^\pm_{\nu}$ where
$\tau^{\pm}_\nu=(\pm i{\vf \tau},1)$ and ${\vf \tau}$ is the
3-vector of Pauli matrices, $p^\pm=p^\nu\tau^\pm_{\nu}$ and $U$ is
a matrix of rotations in the colour space. Surely, we have the
similar relations for the right hand components
$$(\psi^{\dagger
R}_f~R_f~\psi_f^{R})= \int \frat{dp_f
dq_f}{(2\pi)^8}~\psi^{\dagger R}_{f\alpha_f i_f}(p_f)~ R_{\alpha_f
i_f}^{\beta_f j_f}(p_f,q_f;\mu)~\psi_f^{R\beta_f j_f}(q_f)$$
with
the kernel as
$$R_{\alpha i}^{\beta j}(p,q;\mu)=T^{ik}(p;\mu)
\epsilon^{k l}~U^{\alpha}_{l}~U^{\dagger\sigma}_{\beta}
\epsilon^{\sigma n}T^{+}_{n j}(q;-\mu)~,$$ where
$T(p;\mu)=(p+i\mu)^{+}~h^{-}(p;\mu)~,~~
T^+(p;-\mu)=\stackrel{*}h\!\!{^{+}}(p;-\mu)(p+i\mu)^{-}$.
As far
as the vector-function ${\vf h}(p)$ is spanned on the vector ${\vf
p}$ only the components of matrices $(p+i\mu)^{\pm}$ and
$h^{\mp}(p;\mu)$ are permutable. As a result it is easy to
understand the validity of the following identities
$$T(p;\mu)=S^{+}(p;-\mu)~,~~T^+(p;-\mu)=S(p;\mu)~.$$
In what
follows we omit the $\mu$-dependence of matrices $S,T,~S^+,T^+$
because the chemical potential enters the matrices $S^+,T$ always
with the positive sign only and the matrices $S,T^+$ with the
negative one.

After averaging over the colour orientation the leading term of
the $N_c^{-1}$ expansion of four-fermion interaction contribution
looks like the following
\begin{eqnarray}
\left\langle \left(\psi_1^{\dagger L}~L_{1}~\psi_1^{L}\right)~
\left(\psi_2^{\dagger L}~L_{2}~\psi_2^{L}\right)\right\rangle_U&=&
\left(\psi_1^{\dagger L}(p_1)~S(p_1)~
S^{+}(q_1)~\psi_1^{L}(q_1)\right) \left(\psi_2^{\dagger
L}(p_2)~S(p_2)~ S^{+}(q_2)~\psi_2^{L}(q_2)\right)-\nonumber\\
&-&\left(\psi_1^{\dagger L}(p_1)~S(p_1)~
S^{+}(q_2)~\psi_1^{L}(q_2)\right) \left(\psi_2^{\dagger
L}(p_2)~S(p_2)~ S^{+}(q_1)~\psi_1^{L}(q_1)\right)~,\nonumber
\end{eqnarray}
(and, of course, a similar expression for the right hand chiral
component). Eventually, the interaction term may be rewritten as
\begin{eqnarray}
\label{2} {\cal L}_{int}&=&\frat{i\lambda}{4}\left(\psi_f^{\dagger
L}(p_1)~S(p_1)~S^{+}(q_1)~
(\tau^{+}_{a})_{ff'}~\psi_{f'}^{L}(q_1)\right)
\left(\psi_g^{\dagger L}(p_2)~S(p_2)~S^{+}(q_2)~
(\tau^{+}_{a})_{gg'}~\psi_{g'}^{L}(q_2)\right)-\nonumber\\ [-.2cm]
\\ [-.25cm]
&-&\frat{i\lambda}{4}\left(\psi_f^{\dagger
L}(p_1)~T(p_1)~T^{+}(q_1)~
(\tau^{+}_{a})_{ff'}~\psi_{f'}^{L}(q_1)\right)
\left(\psi_g^{\dagger L}(p_2)~T(p_2)~T^{+}(q_2)~
(\tau^{+}_{a})_{gg'}~\psi_{g'}^{L}(q_2)\right)~.\nonumber
\end{eqnarray}
Performing the auxiliary integration over the bosonic fields $L_a$
ш $R_a$ the four-fermion interaction can be transformed to the
Gauss-type integral (analogously for right hand chiral fields and
scalar field $R_a$) $$\frat{\lambda}{4}\left(\psi^{\dagger L}~S
S^{+}~\tau^{+}_{a}~\psi^{L}\right) \left(\psi^{\dagger L}~S
S^{+}~\tau^{+}_{a}~\psi^{L}\right)\to -\lambda \left(\psi^{\dagger
L}~S S^{+}~\tau^{+}_{a}~\psi^{L}\right)~L_a-\lambda L^2_{a}~,$$
which is quite convenient to obtain the effective Lagrangian in
the terms of hadronic degrees of freedom
\begin{eqnarray}
&&\hat L=(1+\sigma+\eta)~U~V~,~~~\hat
R=(1+\sigma-\eta)~V~U^+~,\nonumber\\
&&U=e^{i~\pi^a\tau^a}~,~~~~~~~~~~~~~~~~V=e^{i~\sigma^a\tau^a}~.\nonumber
\end{eqnarray}
Then the corresponding Lagrangian density incorporating the
interaction of quarks and hadrons is
\begin{eqnarray}
\label{3} {\cal
L}^{'}_{int}=&-&i\lambda~\psi^{\dagger}(p)~\left[S(p)~S^{+}(q)~e^{i(p-q)x}~
\left(1+\sigma(x)+\eta(x)\right)~U(x)~V(x)~P_+-\right.\nonumber\\[-.2cm]
\\ [-.25cm]
&-&\left. T(p)~T^{+}(q)~e^{i(p-q)x}~
\left(1+\sigma(x)-\eta(x)\right)~V(x)~U^+(x)~P_-\right]~\psi(q)~,\nonumber
\end{eqnarray}
and calculating two well-known diagrams with two external hadron
lines one may extract the necessary hadronic correlators according
to the corresponding effective Lagrangian \cite{5}
\begin{eqnarray}
\label{4} R_{\pi_a}(p)&=&4N_c \sum\!\!\!\!\!\!\!\int
\frat{dk}{(2\pi)^4}~\frat{M^2(k)}{(k+i\mu)^2+M^2(k)}-\nonumber\\
&-&4N_c \sum\!\!\!\!\!\!\!\int \frat{(dk_1
dk_2)}{(2\pi)^4}~\frat{[(k_1+i\mu)(k_2+i\mu)+M_1 M_2]~M(k_1,k_2)
M(k_2,k_1)}
{\left[(k_1+i\mu)^2+M_{1}^2\right]\left[(k_2+i\mu)^2+M_{2}^2\right]}~,\nonumber\\
R_{\sigma}(p)&=&n\bar\rho^4- 4N_c \sum\!\!\!\!\!\!\!\int
\frat{(dk_1 dk_2)}{(2\pi)^4}~\frat{[(k_1+i\mu)(k_2+i\mu)-M_1
M_2]~M(k_1,k_2) M(k_2,k_1)}
{\left[(k_1+i\mu)^2+M_{1}^2\right]\left[(k_2+i\mu)^2+M_{2}^2\right]}~,\nonumber\\[-.2cm]
\\ [-.25cm]
R_{\eta}(p)&=&n\bar\rho^4+ 4N_c \sum\!\!\!\!\!\!\!\int \frat{(dk_1
dk_2)}{(2\pi)^4}~\frat{[(k_1+i\mu)(k_2+i\mu)+M_1 M_2]~M(k_1,k_2)
M(k_2,k_1)}
{\left[(k_1+i\mu)^2+M_{1}^2\right]\left[(k_2+i\mu)^2+M_{2}^2\right]}~,\nonumber\\
R_{\sigma_a}(p)&=&4N_c \sum\!\!\!\!\!\!\!\int
\frat{dk}{(2\pi)^4}~\frat{M^2(k)}{(k+i\mu)^2+M^2(k)}+\nonumber\\
&+&4N_c \sum\!\!\!\!\!\!\!\int \frat{(dk_1
dk_2)}{(2\pi)^4}~\frat{[(k_1+i\mu)(k_2+i\mu)-M_1 M_2]~M(k_1,k_2)
M(k_2,k_1)}
{\left[(k_1+i\mu)^2+M_{1}^2\right]\left[(k_2+i\mu)^2+M_{2}^2\right]}~.\nonumber
\end{eqnarray}
The momentum integration here includes the $\delta$-function
$(dk_1 dk_2)=dk_1 dk_2~\delta(k_1-k_2-p)$ carrying the external
momentum $p$. The vertex function $M(k_1,k_2)=\lambda S(k_1)
S^+(k_2)=\lambda v(k_1,k_2)$ has the following structure
\begin{equation}
\label{vertx} v(k_1,k_2)=A+iB~\frat{ {\vf k}_1~{\vf \tau}} {|{\vf
k}_1|}+iC~\frat{ {\vf k}_2~{\vf \tau}}{|{\vf k}_2|} +iD~\frat{
{\vf k}_1\times{\vf k}_2~{\vf \tau}}{|{\vf k}_1|~|{\vf k}_2|}~,
\end{equation}
with the functions $$A=\left[k_1 h+(k_1+i\mu)h_4\right]\left[k_2
g+(k_2+i\mu)g_4\right]+
\left[(k_1+i\mu)h-k_1h_4\right]\left[(k_2+i\mu)g-k_2g_4\right]~
\frat{ ({\vf k}_1{\vf k}_2)}{|{\vf k}_1|~|{\vf k}_2|}~,$$
$$B=[(k_1+i\mu)h-k_1h_4][k_2 g+(k_2+i\mu)g_4]~,$$
$$C=-[k_1h+(k_1+i\mu)h_4][(k_2+i\mu)g-k_2g_4]~,$$
$$D=~[(k_1+i\mu)h-k_1h_4][(k_2+i\mu)g-k_2g_4],$$ where
$h=h(k_1;\mu)$, $h_4=h_4(k_1;\mu)$, $g=h(k_2;\mu)$,
$g_4=h_4(k_2;\mu)$, and in the shortened form one should imply
$M_1=M(k_1,k_1)$, $M_2=M(k_2,k_2)$. The product of matrices
$v(k_1,k_2)v(k_2,k_1)$ is spanned on the unit matrix because
$$v(k_2,k_1)=A-iB~\frat{ {\vf k}_1~{\vf \tau}} {|{\vf
k}_1|}-iC~\frat{ {\vf k}_2~{\vf \tau}}{|{\vf k}_2|} -iD~\frat{
{\vf k}_1\times{\vf k}_2~{\vf \tau}}{|{\vf k}_1|~|{\vf k}_2|}~,$$
and denotion $M(k_1,k_2) M(k_2,k_1)$ means the corresponding
scalar to be singled out. Undoubtedly, all expressions of
Eq.(\ref{4}) are approximate and going to improve them one
remembers the matrices $S(k_1) S^+(k_2)$ (and $T(k_1) T^+(k_2)$ as
well) should be linked with the corresponding quark propagators in
between (the vertices might be interchanged only when the momentum
equality $k_1=k_2$ is valid) which finally leads to the additional
terms of the following form $$Tr (A +i {\vf N} {\vf
\Sigma})\left\{\stackrel{ 1}\gamma_5\right\}(\hat k_1+i\hat\mu +i
M_1) (A -i {\vf N} {\vf \Sigma})\left\{\stackrel{
1}\gamma_5\right\}(\hat k_2+i\hat\mu +i M_2)~,$$ where ${\vf
N}=B~\frat{ {\vf k}_1}{|{\vf k}_1|}+C~\frat{ {\vf k}_2}{|{\vf
k}_2|} +D~\frat{ {\vf k}_1\times{\vf k}_2}{|{\vf k}_1|~|{\vf
k}_2|}$. However, numerical analysis performed shows the
additional terms proportional to the corresponding traces of
matrices of the spin type
$\Sigma_i=\frat{i}{2}\varepsilon_{ijk}~\gamma_j \gamma_k$ are
negligible.

\begin{figure*}[!tbh]
\begin{center}
\includegraphics[width=0.75\textwidth]{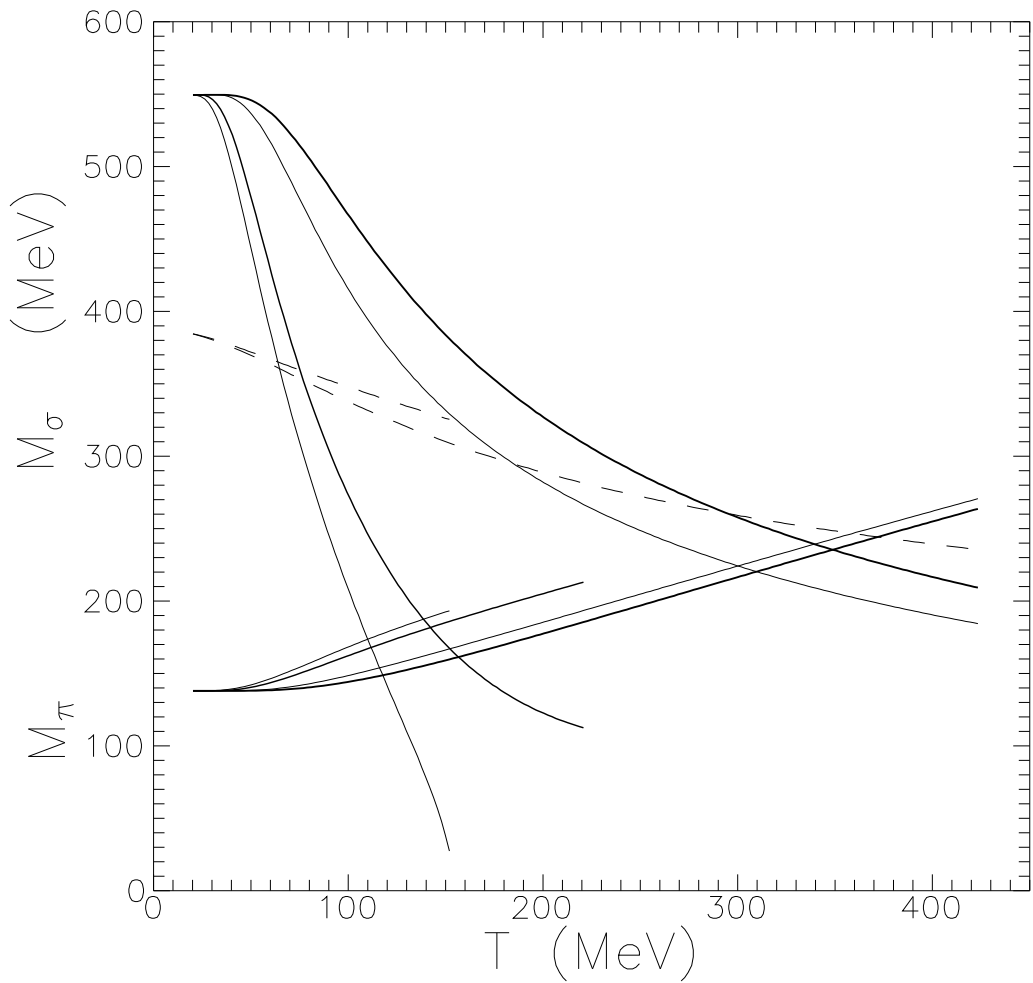}
\end{center}
  \vspace{-7mm}
 \caption{The masses of $\sigma$- (upper solid lines) and $\pi$- mesons (lower solid lines) as the
temperature functions for four values of the chemical potential
$\mu=100, 160, 240, 260$ MeV, the upper line corresponds to
$\mu=100$ MeV. Two dashed lines show the behaviour of dynamical
quark mass corresponding to two extreme values of chemical
potential (upper line corresponds to larger value.}
\label{fig1}
\end{figure*}

Now going to analyse the thermal and finte $\mu$ evolution of
quark mass and quark condensate we would like to emphasize that we
do not expect our approach to be valid in the chiral limit but
reasonably reliable in the precritical region. As usual the saddle
point is determined by the solution of the following equation
\begin{equation}
\label{5} 4N_c \sum\!\!\!\!\!\!\!\int
\frat{dk}{(2\pi)^4}~\frat{M^2(k)}{(k+i\mu)^2+M^2(k)}=n~.
\end{equation}
At zero temperature and particular value of the chemical potential
($\mu_c\simeq 350$ MeV) there appears the singularity in the
denominator of Eq.(\ref{5}) which is rooted in the solution of
equation $${\vf k}^2-\mu^2+M(|{\vf k}|)=0~$$ and quarks are going
to fill the Fermi sphere in. If we do not calculate the imaginary
part of the correlators (\ref{4}) the corresponding integrals
might be calculated as the principal values only. Obviously the
accurate calculation foresees an analysis of all poles and cuts in
the imaginary $p_4$-plane. However, for the verteces generated by
instantons this task, unfortunately, is not feasible today because
of the fairly complicated analytical structure of the zero modes
in the complex plane (see, for example, \cite{11} where
unrealistic (low) thresholds of chiral symmetry restoration have
been received for the zero mode forms considered).

\begin{figure*}[!tbh]
\begin{center}
\includegraphics[width=0.75\textwidth]{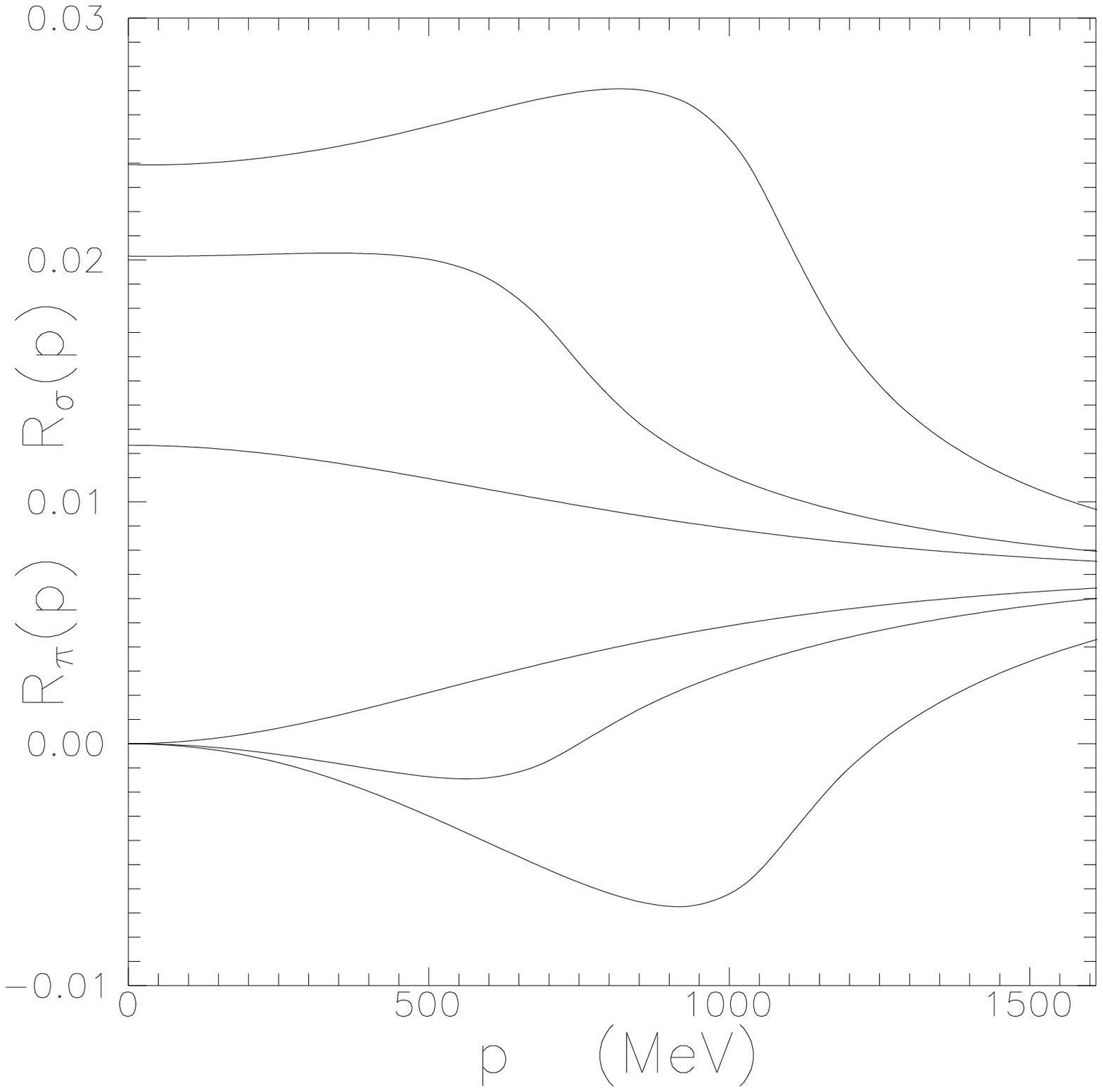}
\end{center}
  \vspace{-7mm}
 \caption{ Correlation functions $R_{\sigma_a}(p)$ (upper solid lines) and $R_\pi(p)$-mesons
(lower solid lines)as the momentum $|{\vf p}|$ (MeV) functions for three magnitudes of chemical
potentials $\mu=0, 510, 680$ MeV.}
  \label{fig2}
\end{figure*}

The masses of $\sigma$- and $\pi$-mesons are extracted from hadron
correlator expansions at small values of momenta (\ref{4})
$R_\pi(p)=\beta_\pi p^2+\dots$,
$R_\sigma(p)=\alpha_\sigma+\beta_\sigma p^2+\dots$ and
$M^{2}_\sigma=\frat{\alpha_\sigma}{\beta_\sigma}$ but the
$\pi$-meson mass is given by the Gell-Mann-Oakes-Renner relation.
In our paper the coefficients $\alpha$ and $\beta$ are numerically
calculated and the precision which we are able to provide for the
meson characteristics is at the level of 10\% only (it is exracted
from the result dependence on the spacing of numerical
differentiation). It looks quite satisfactory for the qualitative
appraisals which here we intend to. The typical thermal behaviours
of $M_\sigma$ and $M_\pi$ are depicted in Fig.1 for the several
values of chemical potentials $\mu= 100, 160, 240, 260$ MeV (a
polarization was taken trivial $|{\vf p}|=0$ while calculating)
and the upper curve there corresponds to $\mu=100$ MeV. The mass
of $\sigma$-meson diminishes because of $\alpha$ decreasing and
$\beta$ increasing at the same time. The distinctive feature of
our approximate calculation is an absence of characteristic
structure having usually observed at the crossing of the $\sigma$-
and $\pi$-meson curves while chiral symmetry is restored
\cite{13}. Two curves of the dynamical mass behaviour
$M(k_4(0);\mu)$ for $\mu=100$ MeV (lower dashed curve) and
$\mu=260$ MeV (upper dashed curve) are shown on the same plot
demonstrating a slow decrease with the temperature increasing. It
is interesting to notice the $\sigma$-meson mass is in the domain
of $\frat{M_\sigma}{M}$ expansion validity with $\mu$ and $T$
increasing which allows us to hope that the results of analysis
with an effective Lagrangian proposed might be taken not only as
qualitative ones. It is well seen that with $\mu$ increasing the
curves describing the $\sigma$-meson mass become steeper and
steeper and around $\mu_c\simeq 350$ MeV where the process of
filling in the Fermi sphere starts $\sigma$-meson is simply
degenerated. Moreover the hadronic correlator
$R_\sigma(p)=\alpha_\sigma+\beta_\sigma p^2+\dots$ develops almost
constant behaviour at $\beta_\sigma \to 0$. With $\mu$ further
increasing the correlation function of $\pi_a$-meson becomes
degenerate but its chiral partner $\sigma_a$-meson appears to be
the 'physical' one. Fig.2 shows the correlation functions of
$R_{\sigma_a}(p)$ and $R_\pi(p)$ mesons as the functions of of
momentum $|{\vf p}|$ (╠¤┬) for three chemical potential values
$\mu= 0, 510, 680$ MeV.

\begin{figure*}[!tbh]
\begin{center}
\includegraphics[width=0.75\textwidth]{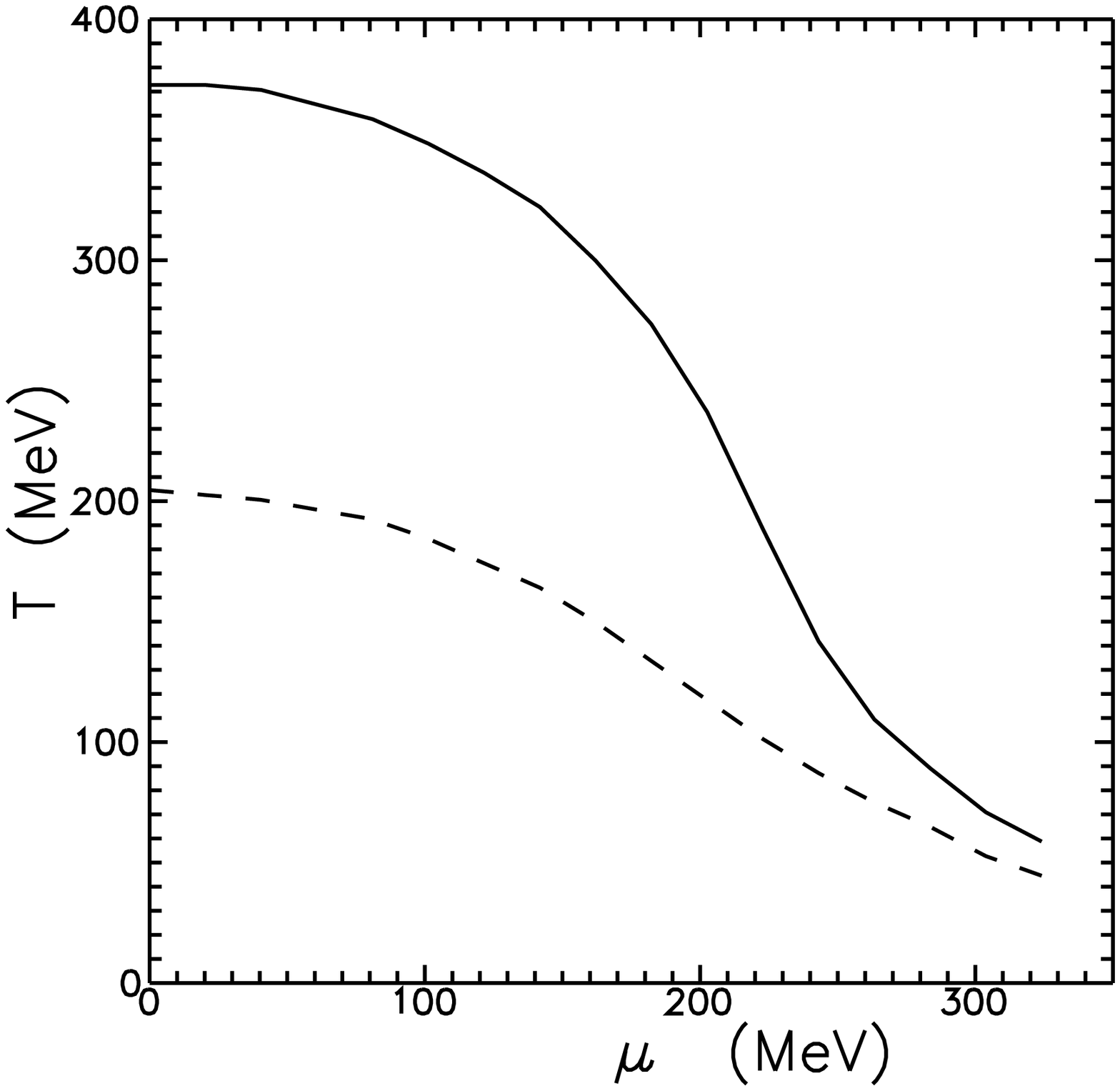}
\end{center}
  \vspace{-7mm}
 \caption{The curve of $M_\sigma=2 M_\pi$ (dashed line) and the curve of $M_\sigma= M_\pi$
(solid line).}
  \label{5fig}
\end{figure*}

The curve where $M_\sigma=2 M_\pi$ is depicted (dashed line) in
Fig.3 together with the line $M_\sigma= M_\pi$ (solid curve).
Unfortunately, more or less reliable theoretical predictions for
their locations on the ($\mu$ -- $T$)-plane do not exist nowadays.
For example, in the Nambu--Jona-Lasinio model these curves have
the intersection point with the $\mu=0$-axis around rather 'high'
critical temperature and their changes while the perturbation
expansion in the saturating configuration is applied is of special
interest. In the model developed, the curves should move down on
the plot and if $\sigma$-meson appears beyond the $M_\sigma=2
M_\pi$ curve the decay channel $\sigma \to 2\pi$ becomes
impossible which favours its experimental identification. In
general we could conclude that our results as to behaviour of
$\sigma$- and $\pi$- mesons on the $\mu -- T$-plane are in
reasonable agreement with the results obtained in other papers
(see, for example, \cite{14}) although one distinction is obvious.
It concerns the quark mass behaviour (see, the respective dashed
lines in Fig.1). Aiming to clarify this point we have studied the
generating functional at zero temperature in the mean field
approximation as it was formulated in \cite{6}. Its considerable
advantage comes from the possibility to calculate the quark
condensate (dynamical quark mass) without appealing the saddle
point equation. However, in this exercise the zero mode
approximation leads to the unphysical increase of quark condensate
when the chemical potential value approaches dynamical quark mass
magnitude. It seems this result dictates a necessity to take into
account the terms neglected even though partly (remember, we have
made use the approximation for zero modes) and to investigate the
role of non-zero modes. But we believe studying the mechanism of
filling the Fermi sphere in gives more illumination to the origin
of chiral symmetry restoration in this case.

In our previous paper \cite{15} we were analysing the possible
mechanism of mixing $\sigma$-meson and the phonon-like excitations
of instanton liquid and demonstrated that notwithstanding the
seeming smallness of coupling constant (small change of dynamical
quark mass) the effects of mixing could be powerful enough. The
specific feature of these effects while depending on $\mu$ and $T$
is the alternation of the heavy and light component roles.
Initially $\sigma$-meson is the heavier component but gn heating
(or compressing) this field loses its mass quickly, as was
demonstrated, and at reaching a certain threshold magnitude the
phonon-like excitations play the role of heavier components.

We are not preoccupied with the claim for the accurate quantative estimates of particular masses or widths
because we realize the approximate character of perturbative scheme proposed. Nevertheless, we believe
this approach could be useful to clarify the possibility of continuing the estimates obtained
to the region of large chemical potentials (larger than dynamical quark mass) and the limits of
applicability of zero mode approximation.

The authors are grateful to A.E. Dorokhov, S.B. Gerasimov,
A.E.Radzhabov and M.K.Volkov for interesting discussions and
constructive criticism. Several discussions with T. Hatsuda were
quite illuminating. Our work was supported by the Grants
INTAS-04-84-398, NATO PDD(CP)-NUKR980668, RFFR 05-0217695 and
and by the special program of the Ministry of Education and Science of the
Russian Federation (grant RNP.2.1.1.5409).

%\newpage

\end{document}